\documentclass[journal=jctcce,manuscript=article,layout=twocolumn]{achemso}
\setkeys{acs}{maxauthors=0}

\usepackage{amsmath}
\usepackage{amssymb}
\usepackage{xcolor}
\usepackage{bm}  
\usepackage{bbm} 
\usepackage{graphicx} 
\usepackage[explicit]{titlesec}
\usepackage{multirow}
\usepackage{float}
\usepackage{xr} 
\usepackage{booktabs}
\usepackage{array}
\usepackage{caption}
\usepackage[section]{placeins}

\definecolor{mygreen}{RGB}{0,112,63} 
\titleformat{\section}{\normalfont\sffamily\bfseries}{{\color{mygreen} $\blacksquare$}}{1em}{{\color{mygreen} \MakeUppercase{#1}}}
\titleformat{\subsection}[runin]{\normalfont\sffamily\bfseries}{}{1em}{#1.}[]
\makeatletter 
\newlength{\figrulesep} 
\makeatother

\author{Thomas Markovich}
\affiliation{Department of Chemistry and Chemical Biology, Harvard University, Cambridge MA}
\author{Martin A. Blood-Forsythe}
\affiliation{Department of Chemistry and Chemical Biology, Harvard University, Cambridge MA}
\author{Dmitrij Rappoport}
\affiliation{Department of Chemistry and Chemical Biology, Harvard University, Cambridge MA}
\author{Dasol Kim}
\affiliation{Department of Chemistry and Chemical Biology, Harvard University, Cambridge MA}
\alsoaffiliation{Department of Chemistry, Mount Holyoke College, South Hadley MA}
\author{Al\'{a}n Aspuru-Guzik}
\email{aspuru@chemistry.harvard.edu}
\affiliation{Department of Chemistry and Chemical Biology, Harvard University, Cambridge MA}

\title{Calibration of the Many-Body Dispersion Range-Separation Parameter}

\begin{document}


\begin{abstract}  
Recent work has shown that a fully many-body treatment of noncovalent interactions, such as that given by the method of many-body dispersion (MBD), is vital to accurately modeling the structure and energetics of many molecular systems with density functional theory (DFT). To avoid double counting the correlation contributions of DFT and the MBD correction, a single-parameter range-separation scheme is typically employed. Coupling the MBD correction to a given exchange-correlation functional therefore requires calibrating the range-separation parameter. We perform this calibration for 24 popular DFT functionals by optimizing against the S66$\times$8 benchmark set. Additionally, we report a linear equation that predicts near optimal range-separation parameters, dependent only on the class of the exchange functional and the value of the gradient enhancement factor. When a calibrated MBD correction is employed, most of the exchange-correlation functionals considered are capable of achieving agreement with CCSD(T)/CBS interaction energies in the S66$\times$8 set to better than 1~kcal/mol mean absolute error.
\end{abstract}

\section{Introduction}
Noncovalent interactions, ranging from hydrogen bonding to weak van der Waals (vdW) forces, present significant challenges to quantum chemical modeling.\cite{cohen_challenges_2012} The difficulty arises due to their many-body quantum nature and small energy scales (typically 5-10~kcal/mol for hydrogen bonds, 1-2~kcal/mol for van der Waals forces).\cite{stone_theory_2013} As a result, while approximate density functional theory (DFT) can often predict the thermochemistry of organic compounds dominated by covalent interactions to within 1~kcal/mol (``chemical accuracy'')\cite{cohen_challenges_2012,becke_perspective_2014}, accurate prediction of noncovalent interactions within DFT remains a challenging task.\cite{riley_2010,marom_dispersion_2011,grimme_review_2011,klimes_2012,dobson_review_2012,berland_review_2015,reilly_van_2015,grimme_chem_rev_2016} At the same time, these noncovalent interactions play an important role in the  energetics, structure, and function of a wide range of systems,\cite{stone_theory_2013} including intermolecular interactions between small molecules\cite{ambrosetti_mbdrsscs_2014,blood-forsythe_analytical_2016}, intramolecular interactions in biomolecules\cite{tkatchenko_unraveling_2011}, and various properties of condensed phase systems such as, molecular crystals\cite{oterodelaroza_jcp_2012,reilly_vibrations_2013,reilly_jpcl_2013,marom_2013,reilly_prl_2014,kronik_2014}, and vdW layered materials\cite{shtogun_many-body_2010,bjorkman_prl_2012}.

The accurate treatment of noncovalent interactions within the framework of DFT is vital for reaching the goal of chemical accuracy for an ever-larger set of experimentally relevant systems. Many different methodologies have been developed to approximately treat noncovalent interactions within the DFT framework\cite{klimes_2012,grimme_review_2011,berland_review_2015,grimme_chem_rev_2016}. Following the classification scheme of \citeauthor{klimes_2012}  (Ref.~\citenum{klimes_2012}), these differing approaches can be categorized into four levels of approximation: I, II, III, and IV. Level I methods involve no explicit correction for noncovalent interactions. Instead, they are paramaterized to reproduce experimental data but provide incorrect asymptotics. Level II methods (generally termed ``DFT-D'') correct DFT functionals with pairwise-additive corrections employing empirical parameters that minimize average errors across large calibration sets. While extremely computationally efficient and enormously popular, DFT-D methods fail to capture the true many-body nature of dispersion, such as the $N$-body dipole terms\cite{axilrod_jcp_1943,muto_jpmsj_1943}, electrodynamic response screening effects\cite{tkatchenko_prl_2012,dobson_review_2012,reilly_chemsci_2015}, and the non-additivity of the dynamic polarizability\cite{wagner_non-additivity_2014,reilly_chemsci_2015}. Indeed, examples of the failings of pairwise methods are numerous, and include incorrect conformer ordering of polypeptide $\alpha$-helices\cite{tkatchenko_unraveling_2011,rossi_stability_2015} and organic crystal polymorphs\cite{oterodelaroza_jcp_2012,reilly_jpcl_2013,reilly_prl_2014}, and poor characterization of the cohesive properties of molecular crystals\cite{shtogun_many-body_2010,bjorkman_prl_2012,distasio_condensed_2014,reilly_vibrations_2013}, among others. Level III consists of non-local kernel-based methods such as vdW-DF\cite{vdW-DF,vdW-DF2} and VV10\cite{VV10}. While these methods get the asymptotic behavior correct, they are unable to capture the presence of dielectrics and still rely on subtle pairwise dipole approximations (see Ref.~\cite{dobson_review_2012} for an explanation of how pairwise approximations appear in non-local kernel methods). Level IV methods treat the pairwise, beyond pairwise, and screening interactions on equal footing, thereby allowing an accurate description of a wider range of dispersively dominated systems with a high degree of confidence. The many-body dispersion (MBD) method~\cite{tkatchenko_prl_2012,ambrosetti_mbdrsscs_2014} discussed herein is one of the most computationally efficient methods at this level of approximation.

The MBD method treats the beyond-pairwise nature of dispersion through a model Hamiltonian that models each atom as a quantum harmonic oscillator, which is fully coupled to other atoms by a screened dipole interaction\cite{tkatchenko_prl_2012,distasio_pnas_2012,distasio_condensed_2014,tkatchenko_jcp_2013,ambrosetti_mbdrsscs_2014}. The long-range correlation energy is then calculated from the difference between the zero-point energy of the coupled oscillator system and that of the  uncoupled oscillators. The procedure of diagonalizing the Hamiltonian of these coupled oscillators is equivalent (through the adiabatic-connection fluctuation-dissipation theorem) to the random-phase approximation (RPA) correlation energy in the dipole limit.~\cite{tkatchenko_jcp_2013,ambrosetti_mbdrsscs_2014} 

To date, the MBD model has only been applied as a correction to the PBE~\cite{PBE_ref1,PBE_ref2}, PBE0~\cite{PBE0_ref1,PBE0_ref2,PBE0_ref3}, and HSE~\cite{HSE_ref1,HSE_ref2} exchange-correlation functionals.\cite{ambrosetti_mbdrsscs_2014,maurer_many-body_2015} While these functionals have proved generally successful for a broad range of systems, many functionals exist that have been specifically designed to perform well for specific classes of molecules\cite{MOHLYP, mohlyp2_ref}, or specific chemical applications such as thermochemistry\cite{thermochem_1, thermochem_2}, kinetics\cite{thermochem_1, kinetics_1, thermochem_2}, or spectroscopy\cite{spectroscopy_1}. Like other dispersion corrections, the MBD model employs a damping function to avoid double counting the correlation energy at short-range where the exchange-correlation functional is expected to give a good description of correlation. The MBD range-separation parameter, $\beta$, which determines the range of the damping function must be tuned carefully for each functional because the short-range behavior of the exchange-correlation potential (particularly the repulsive exchange wall) will dramatically impact the shape of a binding curve. In this work we extend the applicability of the MBD model to other functionals by performing this calibration. We present the optimal range-separation parameter for MBD, as well as benchmark results, for 24 popular exchange-correlation functionals. 

\section{The MBD Method}
The following is a brief introduction to the theory and notation of the MBD model. All equations are given in Hartree atomic units ($\hbar$ = $m_e$ = e = 1) with tensor  quantities denoted by bold typeface.  We refer the reader to Ref.~\citenum{blood-forsythe_analytical_2016} for a more detailed derivation and discussion of the equations presented below. The canonical London approximation for dispersion interactions describes small instantaneous oscillations of the charge density that induce local dipoles. As these dipoles oscillate about their equilibrium positions, they interact through the Coulomb interaction. With this picture in mind, modeling these density oscillations with quantum harmonic oscillators (QHO) provides a physically intuitive description. One QHO is centered on each atomic nucleus, parameterized with a frequency-dependent scalar dipole polarizability, $\alpha_a(\mathbbm{i} \omega)$, and a characteristic frequency, $\omega_a$, where $a$ is an atomic/nuclear index. Here, the polarizability plays the role of the QHO mass and the characteristic frequency is directly related to the strength of the QHO potential. These parameters are determined by reweighting reference atomic polarizabilities by the ratio of the Hirshfeld atomic volume~\cite{hirshfeld_1977} to the free atom volumes for each atom. This captures effects of the local chemical environment without introducing adjustable parameters. This partitioning is in contrast to methods such as DFT-D3, which seek to capture the local chemical environment from the geometry alone by parameterizing coordination numbers. Purely geometric approaches miss important density-dependent changes to the polarizability that can occur from solvation or the presence of external electric fields.

In addition to local changes to the atomic static polarizabilities, electrodynamic screening and the anisotropy of chemical bonds are modeled by solving a Dyson-like self-consistent screening equation to compute screened dynamic polarizabilities, $\overline{\alpha}_a(\mathbbm{i}\omega)$, for each QHO. Using these screened polarizabilities, we compute the corresponding screened characteristic frequency, $\overline{\omega}_a$, for each QHO by numerical evaluation of the Casimir-Polder integral and inversion of a 0/2-order Pad\'{e} approximant~\cite{tang_pr_1968} to the frequency dependence of the scalar dipole polarizability
\begin{equation}
\overline{\omega}_a = \left( \frac{4}{3} \frac{1}{\big[\overline{\alpha}(0)\big]^2} \right) \left( \frac{3}{\pi} \int^{\infty}_0 \big[\overline{\alpha}_a(\mathbbm{i} \omega) \big]^2 d \omega \right).
\end{equation}
The MBD model Hamiltonian for an $N$ atom system is then constructed from these screened polarizabilities and excitation frequencies as
\begin{eqnarray}\label{eq:H}
{\rm H}_{\rm MBD} &=& -\sum_{a=1}^N \frac{1}{2} \nabla^2_{\bm{\mu}_a} + \sum^N_{a=1} \frac{1}{2} \overline{\omega}^2_{a} \bm{\mu}^2_a \\ 
&&+ \sum^N_{a>b} \overline{\omega}_a \overline{\omega}_b \sqrt{\overline{\alpha}_a(0)\, \overline{\alpha}_b(0)} \bm{\mu}_a^\dagger \mathbf{T}_{\rm LR}^{\,ab} \bm{\mu}_b \nonumber ,
\end{eqnarray}
where $\bm{\mu}_a$ is the polarizability-weighted displacement of the $a^{\rm th}$ QHO, and $\mathbf{T}^{ab}_{\rm LR}$ is the long-range component of the dipole-dipole interaction tensor. The first two terms can be identified as the base QHO Hamiltonian in polarization weighted coordinates, while the third term represents two-body coupling. This coupling is calculated from the long-range contribution to the dipole-dipole interaction tensor weighted by a Fermi-type damping function, so the $(i,j)^{\rm th}$ Cartesian component is:
\begin{equation}
\mathrm{T}^{\,ab, \,ij}_{\rm LR} = \left[ \frac{1}{ 1 + \exp\left[-Z_{ab}\right] } \right] \frac{-3 \mathbf{R}_{ab}^i \mathbf{R}_{ab}^j + \left\|\mathbf{R}_{ab}\right\|^2 \delta_{ij} }{ \left\|\mathbf{R}_{ab}\right\|^5 },
\end{equation}
where
\begin{equation}\label{eq:z}
Z_{ab} = 6 \left[ \frac{ \left\|\mathbf{R}_{ab}\right\| }{ \beta \left( \mathcal{R}_a^{\rm vdW} + \mathcal{R}_b^{\rm vdW} \right) } - 1 \right]
\end{equation}
is the ratio between the internuclear distance, $\left\|\mathbf{R}_{ab}\right\|$, and the modified sum of the effective van der Waals (vdW) radii, $\mathcal{R}^{\rm vdW}$,  of atoms $a$ and $b$. The parameter $\beta$ is an adjustable range-separation parameter that determines the length scale at which the damping function acts to ``turn on'' the MBD contribution. To prevent double counting of the correlation energy, the MBD correction should ``turn on'' only after the correlation contribution from the underlying DFT exchange-correlation functional becomes sufficiently small. $\beta$ can be thought of as an inverse length scale. As $\beta \rightarrow 0$, the MBD contribution never turns off, and as $\beta \rightarrow \infty$, the MBD contribution never turns on. While we are free to choose $\beta$, it is typically fit once for a given exchange-correlation functional by minimizing the mean absolute error with respect to highly accurate reference data. This calibration has previously been done for PBE, PBE0, and HSE using as reference the S66$\times$8 benchmark set\cite{rezac_s66x8_2011}, which is comprised of 528 interaction energies of 66 molecular dimers computed with CCSD(T) extrapolated to the complete basis-set limit.  

Diagonalization of the interacting Hamiltonian produces $3N$ polarization normal modes with corresponding eigenfrequencies, $\sqrt{\lambda_p}$.~\footnote{As usual, the eigenvalues $\lambda_p$ are the square of the frequencies, $\sqrt{\lambda_p}$, associated with the normal modes.} The MBD correlation energy is then computed from the difference of two zero point energies, the first obtained from summing these eigenfrequencies and the second from summing the screened characteristic frequencies of the uncoupled oscillators:
\begin{equation}
{\rm E}_{\rm MBD} = \frac{1}{2} \sum^{3N}_{p=1} \sqrt{\lambda_p} - \frac{3}{2} \sum^N_{a=1} \overline{\omega}_a.
\end{equation}
Finally, ${\rm E}_{\rm MBD}$ is applied as an additive correction to the DFT exchange-correlation energy, so the total energy is
\begin{equation} \label{eq:etot}
\rm E_{\rm total}(\beta) = E_{\rm kinetic}+ E_{\rm XC} +  E_{\rm MBD}(\beta) .
\end{equation}
Because the MBD energy is a function of $\beta$, we can straightforwardly minimize the deviation of the DFT+MBD total energy in equation~\eqref{eq:etot} from a reference energy by varying $\beta$. 


\section{Methods}
\subsection{MBD and DFT Energies}
All DFT calculations were carried out in a development version of Octopus~\textsc{v5.0.0}\cite{marques_octopus_2003,castro_octopus_2006,andrade_tddft_2012,andrade_realspace_2015}, in which we have implemented routines to compute the MBD energy as well as analytic nuclear gradients and analytic functional derivatives of the MBD energy. The Octopus package uses of the LibXC\cite{marques_libxc_2012} library of exchange-correlation functionals, which provides access to many more functionals than are currently available in Quantum ESPRESSO\cite{giannozzi_2009}, the other open-source quantum chemistry package that currently contains an implementation of the MBD correction.\cite{blood-forsythe_analytical_2016} As an added benefit, Octopus' use of real-space grids provides an efficient treatment of gas-phase molecules and other isolated systems. 

When MBD is applied non-self-consistently it has no impact on the charge density. As a result, the Hirshfeld volumes do not depend on the range-separation parameter, $\beta$, and may be saved to disk for each monomer and dimer in the calibration set. Thus, we optimized $\beta$ by computing each interaction energy once per exchange-correlation functional and then applied MBD as an {\it a posteriori} correction using a locally developed standalone MBD code that accepts $\beta$ as an input parameter together with the atomic coordinates and Hirshfeld volumes. This scheme avoids unnecessary duplication of a large number of DFT single-point calculations. 

\subsection{Basis Sets}
All calculations were run with Octopus' built in library of Hartwigsen-Goedeker-Hutter (HGH) dual space Gaussian norm-conserving pseudopotentials\cite{goedecker_separable_1996,hartwigsen_relativistic_1998} on spherical atom grids of radius of $6~\rm au$ with a grid spacing of $0.08~\rm au$. We tested these pseudopotentials, grid parameters, and grid shapes extensively to verify that they provided energies that agreed to better than $0.02~\rm kcal/(mol \cdot atom)$  with respect to all-electron Gaussian-type orbital (GTO) complete-basis set extrapolated calculations run in Q-Chem~\textsc{v4.1}\cite{qchem_ref2}. We also explored the energy convergence of molecular single-point energies in Octopus for grid spacings ranging from $0.5$ to $0.02~\rm au$, and observed that grid spacings $\leq 0.08~\rm au$ provided convergence within $0.02~\rm kcal/(mol \cdot atom)$ of the value obtained by a quadratic extrapolation to the complete-basis set limit. 

\subsection{$\bm{\beta}$ Optimization}
We determined the optimal range-separation parameter, $\beta_{\rm opt}$, by computing the MBD energy over a grid of $\beta$ values between $0.35$ and $1.19$, with a spacing of $0.01$. We have found that with $\beta$ values below $0.35$, MBD becomes numerically unstable and yields imaginary frequencies when diagonalizing the interacting Hamiltonian. The optimal $\beta$ value was selected to minimize the mean absolute error with respect to CCSD(T)/CBS interaction energies from the S66$\times$8~\cite{rezac_s66_2011,rezac_s66x8_2011} benchmark set of \citeauthor{rezac_s66x8_2011}.  The S66$\times$8 benchmark includes interaction energies for 66 molecular dimers designed to be representative of the types of noncovalent interactions that commonly occur in organic molecules and biomolecules, including electrostatic (hydrogen bonding) dominated, dispersion dominated (aromatic-aromatic, aromatic-aliphatic, and aliphatic-aliphatic), and mixed character interactions with a variety of bonding motifs for each interaction type. The S66$\times$8 set systematically explores 8 points along the dissociation curves of each complex; by optimizing $\beta$ against a dataset that includes non-equilibrium geometries, we avoid introducing a bias towards equilibrium structures.  
 
To determine the consistency of this optimal $\beta$ value across reference datasets, we also optimized $\beta$ against the revised S22~\cite{jurecka_s22_2006,takatani_s22a_2010} benchmarks of Sherrill and co-workers~\cite{takatani_s22a_2010}. The S22 dataset contains 22 small molecule dimers covering a mix of hydrogen bonded and dispersion bound complexes, although it is somewhat weighted toward nucleic acid-like structures and less balanced in its coverage of interaction motifs than the S66 set~\cite{rezac_s66_2011}. Results of optimizing $\beta$ against the S66 and S22 benchmarks are presented in the Supporting Information.

\begin{figure}[tp!]
\centering
\includegraphics[width=228pt]{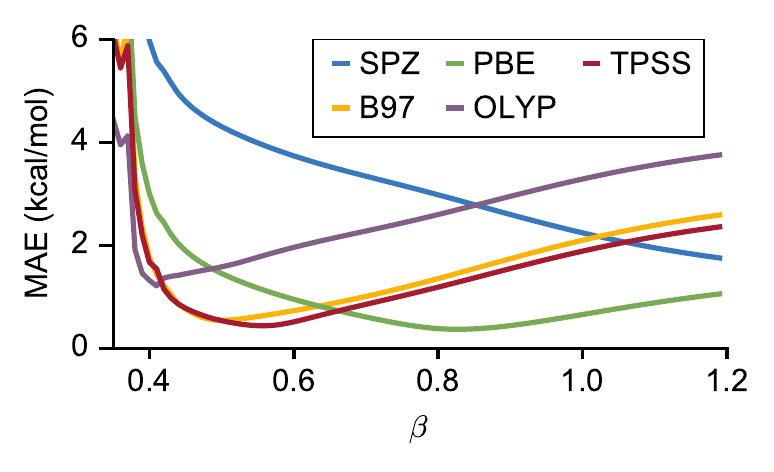}
\caption{The mean absolute error (MAE) in kcal/mol with respect to the S66$\times$8 benchmark set for five separate functionals is presented against the MBD range separation parameter, $\beta$.\label{fig:beta_curves}}
\end{figure}

\section{Results and Discussion}
To facilitate clarity, Table~\ref{tab:fun_ref} of the Supporting Information provides a glossary that details the exchange and correlation functionals associated with each DFT functional abbreviation used in the following discussion of our results.

In Figure~\ref{fig:beta_curves} we present a representative plot of the variation of the mean absolute error (MAE) as $\beta$ is scanned for five different DFT functionals that cover the diversity of exchange present in our 24 functional test suite: SVWN~\cite{slater_exchange_ref1,slater_exchange_ref2,VWN_rpa}, PBE\cite{PBE_ref1,PBE_ref2}, OLYP\cite{optX,C_LYP_ref1,C_LYP_ref2}, TPSS\cite{TPSS_ref1,TPSS_ref2}, and B97\cite{B97}. We observe flat minima for PBE, TPSS, and B97 with a wide range of $\beta$ values resulting in $<1$~kcal/mol error. This behavior is representative of most functionals in the test suite. OLYP displays a narrower minima with an abrupt jump at $\beta=0.42$. Unfortunately, we have no clear explanation for this behavior, though we note that similar discontinuities are observed for a few other functionals at the same $\beta$ value. We have tested our methods extensively and found them to produce no negative eigenvalues in the $\beta$ range of interest. We believe that this may be an artifact that results from coupling oscillators together that are too close to one another. It is well known that LDA functionals exhibit spurious exchange binding that mimics an over-bound dispersive attraction, but is entirely unphysical in its origin; therefore, adding a dispersion correction on top of LDA is expected to degrade the performance of LDA functionals. This is consistent with the behavior of SVWN observed in Figure~\ref{fig:beta_curves}, with the MAE decreasing monotonically as $\beta$ increases. We have confirmed that this monotonic trend continues out to $\beta=10$ for LDA functionals, consistent with $\beta\to\infty$ being optimal, \textit{i.e.} turning off the MBD correction completely. 

\begin{figure*}[tb!]
\centering
\includegraphics[width=457pt]{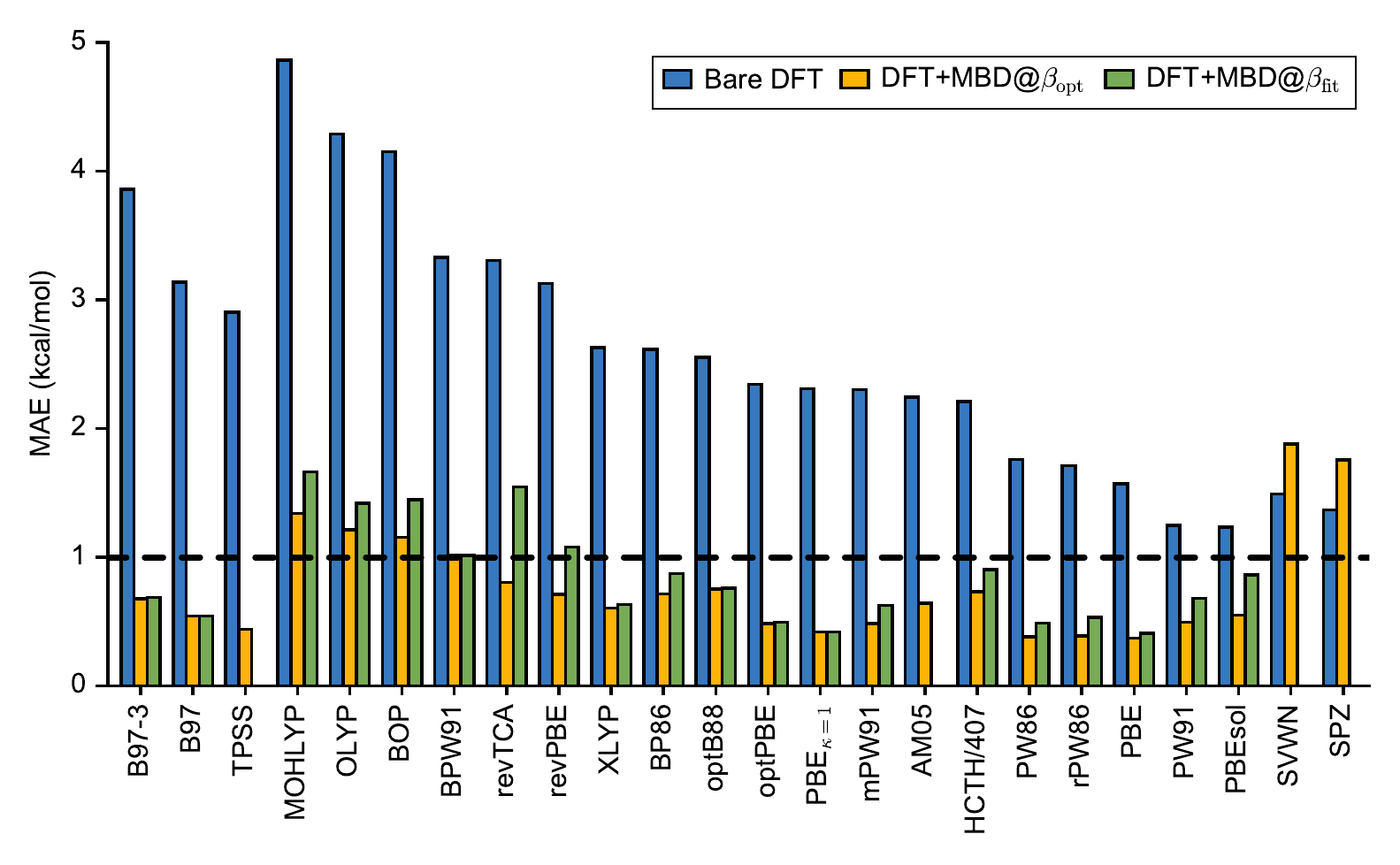}
\caption{Mean absolute error (MAE) in kcal/mol with respect to the S66$\times$8 benchmark set for each functional considered. Bare DFT indicates the MAE of the uncorrected DFT functional. DFT+MBD@$\beta$ indicates the MAE of the MBD corrected functional, with MBD computed using either $\beta_{\rm opt}$ or $\beta_{\rm fit}$. The optimized range-separation parameter ($\beta_{\rm opt}$) is determined by minimizing the MAE; whereas, $\beta_{\rm fit}$ is derived from a linear fit of $\beta_{\rm opt}$ against the exchange functional's gradient enhancement factor. The horizontal dotted line at $1~\rm kcal/mol$ indicates ``chemical accuracy''. 
\label{fig:maebars}}
\end{figure*}

In Figure~\ref{fig:maebars}, we present the MAEs for our full collection of functionals, both with and without the MBD correction. Table~\ref{tab:delta} provides a different view of these results, demonstrating the fractional improvement in MAE that is offered by employing the optimized MBD correction. In the Supporting Information, we present additional tabulations of the MAEs and mean absolute relative errors (MAREs) for the S66$\times$8, S66, and S22 benchmark sets in Tables~\ref{tab:s22mae}-\ref{tab:s66x8mare_broken_down}. We have found that with the exception of the two LDA functionals (SPZ and SVWN), the addition of MBD greatly improves the performance of the functionals considered and most are able to achieve ``chemical accuracy'' when paired with an optimized MBD. 
 
\begin{table*}[tb!]
\centering
\caption{
Fractional improvement ($\Delta$) in the mean absolute error (MAE) with respect to the S66$\times$8 benchmark set~\cite{rezac_s66x8_2011} when using the range-separated MBD correction at the optimal $\beta$ value, indicated by $\beta_{\rm opt}$, is presented for the considered exchange-correlation functionals. $\Delta$ is computed as the ratio of the MAE for the uncorrected functional over that of the MBD corrected functional. For instances in which the MBD correction results in a worse MAE, $\Delta$ is computed as the negative ratio of the MBD corrected MAE over that of the uncorrected functional. In addition to considering the total S66$\times$8 set, we independently optimized $\beta$ on three subgroups of complexes categorized by their dominant bonding motif: hydrogen bonds, dispersion, and other mixed character interactions.
\label{tab:delta}}
\begin{tabular}{ l | cc | cc | cc | cc }
\toprule
\multicolumn{1}{l}{} & \multicolumn{2}{c}{ Total } & \multicolumn{2}{c}{ H-Bonds } & \multicolumn{2}{c}{ Dispersion } & \multicolumn{2}{c}{ Other } \\ 
 \cmidrule(lr){2-3}  \cmidrule(lr){4-5}  \cmidrule(lr){6-7} \cmidrule(lr){8-9}  
 \multicolumn{1}{l}{Functional} &  \multicolumn{1}{c}{$\beta_{\rm opt}$} & \multicolumn{1}{c}{$\Delta$} &   \multicolumn{1}{c}{$\beta_{\rm opt}$} & \multicolumn{1}{c}{$\Delta$} &  \multicolumn{1}{c}{$\beta_{\rm opt}$} & \multicolumn{1}{c}{$\Delta$} &  \multicolumn{1}{c}{$\beta_{\rm opt}$} & \multicolumn{1}{c}{$\Delta$} \\
\midrule
B97-3		&	0.45	&	5.7	&	0.44	&	11	&	0.55	&	4.5	&	0.42	&	6.1	\\
B97			&	0.50	&	5.7	&	0.49	&	8.6	&	0.61	&	4.8	&	0.52	&	7.7	\\
TPSS		&	0.56	&	6.6	&	0.57	&	12	&	0.48	&	6.0	&	0.49	&	8.0	\\
MOHLYP		&	0.40	&	3.6	&	0.45	&	4.5	&	0.39	&	3.8	&	0.39	&	4.7	\\
OLYP		&	0.41	&	3.5	&	0.49	&	5.0	&	0.41	&	4.1	&	0.39	&	4.9	\\
BOP			&	0.42	&	3.6	&	0.52	&	4.7	&	0.40	&	4.4	&	0.40	&	6.0	\\
BPW91		&	0.53	&	3.3	&	0.62	&	3.6	&	0.43	&	6.3	&	0.42	&	5.8	\\
revTCA		&	0.52	&	4.1	&	0.55	&	8.1	&	0.43	&	4.6	&	0.42	&	5.7	\\
revPBE		&	0.54	&	4.4	&	0.59	&	6.6	&	0.43	&	5.7	&	0.44	&	7.2	\\
XLYP		&	0.65	&	4.3	&	0.71	&	7.6	&	0.49	&	5.0	&	0.52	&	6.5	\\
BP86		&	0.64	&	3.6	&	0.75	&	3.1	&	0.47	&	5.5	&	0.53	&	5.0	\\
optB88		&	0.65	&	3.4	&	0.79	&	2.5	&	0.49	&	5.3	&	0.57	&	4.6	\\
optPBE		&	0.67	&	4.8	&	0.75	&	3.7	&	0.59	&	6.7	&	0.62	&	6.6	\\
PBE$_{\kappa=1}$&	0.68&	5.5	&	0.74	&	5.1	&	0.60	&	6.4	&	0.63	&	7.9	\\
mPW91		&	0.68	&	4.7	&	0.77	&	3.9	&	0.57	&	6.8	&	0.64	&	7.1	\\
AM05		&	0.71	&	3.5	&	0.87	&	2.0	&	0.57	&	5.7	&	0.68	&	5.4	\\
HCTC/407	&	0.67	&	3.0	&	0.70	&	2.6	&	0.60	&	3.2	&	0.63	&	3.6	\\
PW86		&	0.80	&	4.6	&	0.93	&	3.4	&	0.74	&	8.2	&	0.76	&	6.3	\\
rPW86		&	0.81	&	4.4	&	0.95	&	3.1	&	0.75	&	8.5	&	0.77	&	6.1	\\
PBE			&	0.83	&	4.2	&	0.99	&	1.9	&	0.77	&	8.3	&	0.82	&	7.6	\\
PW91		&	0.91	&	2.5	&	1.19$^*$&	1.2	&	0.83	&	5.0	&	0.91	&	3.3	\\
PBEsol		&	0.97	&	2.2	&	1.19$^*$&	-1.1	&	0.93	&	6.1	&	0.96	&	4.0	\\
SVWN		&1.19$^*$	&	-1.3&	1.19$^*$&	-1.1	&	1.19$^*$&	-1.7	&	1.19$^*$&	-1.4	\\
SPZ			&1.19$^*$	&	-1.3&	1.19$^*$&	-1.1	&	1.19$^*$&	-1.7	&	1.19$^*$&	-1.4	\\
\bottomrule
\multicolumn{9}{p{4.5in}}{$^*$ Maximum value of $\beta$ considered. LDA functionals are expected to optimize to $\beta \to \infty$.}
\end{tabular}
\end{table*}

In general, the functionals seem to require a larger $\beta_{\rm opt}$ for systems with more hydrogen bonding character than dispersion character. This implies that the underlying functional is providing a better treatment of hydrogen bonding/electrostatic interactions than dispersion. Given that hydrogen bonding is an exchange dominated effect, while dispersion interactions are correlation dominated, it is unsurprising that a long-range correction to the correlation energy is not as beneficial for hydrogen-bonded systems. There are two notable exceptions to this trend, for both B97 and B97-3\cite{B97_3}, which are both global hybrid functionals,  $\beta_{\rm opt}$ for dispersion is larger than $\beta_{\rm opt}$ for hydrogen-bonded systems. It is hard to draw any general conclusions about global hybrids from these results. However, for B97 and B97-3 specifically, this behavior is likely caused by an overly repulsive exchange wall. Octopus' implementation of exact exchange is inefficient enough that a wider survey of local and global hybrid functionals was computationally prohibitive. 

In agreement with previous work, the optimal $\beta$ for PBE was found to be $0.83$\cite{ambrosetti_mbdrsscs_2014} and produced an MAE of  $0.37~\rm kcal/mol$ against the S66$\times$8 set, which is the smallest error in our simulations. We next considered three PBE-like functionals, revPBE\cite{revPBE,revPBE_reply}, PBE$_{\kappa=1}$\cite{PBE_ref1,PBE_ref2,klimes_2010}, and optPBE\cite{PBE_ref1,PBE_ref2,klimes_2010}, which were reparameterized to improve their short-range behavior. PBE$_{\kappa=1}$ and optPBE, were explicitly constructed to allow for the inclusion of noncovalent and long-range corrections such as MBD or D3 without double counting short-range correlation. revPBE, by contrast, was reparameterized to obey the Oxford-Lieb bound on the exchange-correlation energy\cite{revPBE}. In all three cases, we find a smaller $\beta_{\rm opt}$ than that of PBE. This agrees well with the observation that the gradient enhancement factors, $F_x(s)$, for revPBE, PBE$_{\kappa=1}$, and optPBE were designed to be shorter range than PBE. Given that these functionals differ only slightly in the parameters of their exchange interaction, it is unsurprising that their average performances are quite similar, with optPBE, PBE$_{\kappa=1}$, and revPBE having MAEs of $0.49$, $0.42$, and $0.71~\rm kcal/mol$ respectively.

The next functional considered with PBE-like exchange was revTCA\cite{revTCA}, which  is the combination of a revised PBE exchange functional formulated to globally obey the Lieb-Oxford bound, and a correlation functional constructed to give zero correlation energy for hydrogenic atoms. While the uncorrected revTCA functional was among the worst of the functionals tested, yielding an MAE of $3.31~\rm kcal/mol$, revTCA+MBD was able to achieve an excellent MAE of $0.52~\rm kcal/mol$. Previous studies demonstrated that revTCA provides better predictions than PBE for several standard test properties such as barrier heights, atomization energies, and activation enthalpies, and performs comparably to revPBE~\cite{revTCA,tognetti_2010}. The similarity between revTCA and revPBE is reflected in their optimal $\beta$ values: $0.52$ and $0.54~\rm{kcal/mol}$ respectively. Across all binding motifs (see Table~\ref{tab:s66x8mae_broken_down}), we observe that revTCA has a similar optimal $\beta$ value to revPBE, but is slightly shorter range. This correlation in optimal range-separation across binding motifs suggests that the exchange walls described by these two functions is quite similar.

In the same PBE family, PBEsol\cite{PBEsol} was parameterized to perform more similarly to LDA functionals for condensed phase systems. This was achieved largely by damping the gradient enhancement factor at intermediate range to allow for LDA exchange to play a larger role. It is no surprise then that the $\beta_{\rm opt}$ value for PBEsol is larger than that of PBE due to LDA's spurious exchange binding. When the S66$\times$8 set is segmented into binding motif subsets (see Table~\ref{tab:s66x8mae_broken_down}), we see that like LDA, uncorrected PBEsol sufficiently binds hydrogen-bonded complexes and thus the addition of MBD to PBEsol for these systems incurs  additional error; this is reflected by the negative $\Delta$ value in Table~\ref{tab:delta}. It is notable that PBEsol+MBD's performance on hydrogen-bonded systems is the only observed instance where MBD increases the MAE of a non-LDA functional. 

Previous authors have found that the non-empirical PW86\cite{PW86} exchange functional provides the best description of the repulsive part of the exchange potential relevant to van der Waals (vdW) interactions at short range\cite{VV10}. Because PW86 exchange avoids spurious exchange binding, it is therefore a good candidate for the foundation of a vdW corrected functional\cite{vdW-DF2}. Furthermore, PW86 has a revised variant, rPW86\cite{PW86,vdW-DF2}, which more correctly obeys the Lieb-Oxford bound and was constructed to offer improved performance with vdW corrections\cite{VV10}. Both PW86 and rPW86 achieve some of the lowest errors against the S66$\times$8 set, with MAEs of $0.38$ and $0.39~\rm kcal/mol$, respectively. We found that the optimized range-separation parameters differed by $0.01$, with PW86 and rPW86 giving $\beta_{\rm opt}=0.80$ and $0.81$, respectively. The similarity of these optimal $\beta$ values to that of PBE is expected, given the similarity in the functional form of their exchange. 

Both PW91\cite{PW91_ref1,PW91_ref2,PW91_ref3} and mPW91\cite{PW91_ref1,PW91_ref2,PW91_ref3,X_mPW91} performed well when the MBD correction was applied. The uncorrected PW91 and mPW91 functionals achieved MAEs of $1.25$ and $2.30~\rm kcal/mol$ respectively, while the MBD correction reduced these errors to an impressive $0.50$ and $0.49~\rm kcal/mol$ respectively. These errors being within $0.15~\rm kcal/mol$ of the PBE+MBD results is perhaps unsurprising given the fact that PBE exchange was derived to be energetically similar to PW91 exchange. We observe that uncorrected PW91 performs well on the hydrogen-bonded subset and benefits by only a $0.1~\rm kcal/mol$ reduction in MAE with the addition of MBD. By contrast, mPW91 provides a considerably worse description of hydrogen bonding and requires MBD to turn on much quicker, as indicated by $\beta_{\rm opt}=0.77$. Generally, the PW91 family of functionals performs quite well when corrected with MBD.

We found that of the functionals tested, those based on B88 exchange\cite{B88}, including BOP\cite{B88,OP}, BPW91\cite{B88,PW91_ref1,PW91_ref2,PW91_ref3}, BP86\cite{B88,P86}, and optB88\cite{PBE_ref1,PBE_ref2,klimes_2010} were among the worst functionals when corrected with MBD. While optB88 and BP86 were able to achieve MAEs $<1~\rm kcal/mol$ with an optimized MBD correction, BPW91 and BOP achieved MAEs of $1.02$ and $1.15~\rm kcal/mol$, respectively. Judging from the optimal $\beta$ values, all functionals based on B88 appear much shorter range than PBE, with BOP being the shortest at $\beta_{\rm opt}=0.42$ and optB88 the longest, with $\beta_{\rm opt}=0.65$. All of the B88-based functionals considered struggle with treating dispersion dominated complexes (see Table~\ref{tab:s66x8mae_broken_down}), while performing significantly better on other binding motifs. In the case of BOP, the poor treatment of the dispersion dominated dimers contributes the most to the overall error. The final functional that we considered with B88 exchange was XLYP\cite{XLYP,C_LYP_ref1,C_LYP_ref2}, which uses a weighted combination of PW91 and B88 exchange. As such, it is unsurprising that the MAE for XLYP falls between that of PW91 and BP86. Indeed, the optimal $\beta$ for XLYP is in between that of BP86 and PW91, lying closer to BP86 due to the $72.2\%$ B88 contribution to XLYP exchange.  

We examined two functionals, OLYP~\cite{optX,C_LYP_ref1,C_LYP_ref2} and MOHLYP~\cite{MOHLYP}, that use Handy and Cohen's OPTX exchange functional~\cite{optX}, and found that they were among the worst performers in the set, even with the MBD correction. The poor performance of these functionals is unsurprising given that OPTX has an overly repulsive short-range exchange potential; this is reflected in the very short range $\beta_{\rm opt}\simeq 0.4$. OLYP attains a MAE of $4.29~\rm kcal/mol$, which is lowered to $1.21~\rm kcal/mol$ with the application of MBD. The MOHLYP functional contains a scaled version of OPTX and obtained a MAE of $4.86~\rm kcal/mol$, which was reduced to $1.34~\rm kcal/mol$ with MBD. Although the MBD correction provides a large fractional improvement of 3.5 and 3.6 for OLYP and MOHLYP respectively, they are still markedly inferior to the other LYP correlation~\cite{C_LYP_ref1,C_LYP_ref2} based functional XLYP. MOHLYP was optimized for transition metal chemistry, emphasizing the role of d-orbital interactions at the cost of $\pi$-delocalized systems, which explains its generally poor performance on the S66$\times$8 set. While this is a justifiable design for treating organometallic systems, this is the wrong set of physical approximations required to appropriately treat the systems in the  S66$\times$8 set. Even still, we find that MBD significantly improves MOHLYP's treatment of dispersion interactions.

One of the most unusual functionals in our test set is the AM05 functional. AM05 was constructed to treat molecules and surfaces on the same footing as bulk solids. While AM05 has been remarkably successful in treating surfaces and solids, it has rarely been applied to molecules in the gas-phase. Achieving an MAE of $0.64~\rm kcal/mol$ when dispersion corrected, AM05 is capable of high accuracy on molecular interaction energies. AM05 is notable as the functional with the largest spread in $\beta_{\rm opt}$ values for the three binding motifs, ranging from 0.57 to 0.87. It is worth noting that the MBD optimization curve is particularly flat for AM05, indicating that the MBD contribution is a robust correction that is less sensitive to the particular form of AM05's exchange potential than with other functionals.

We included two global hybrids in our set of functionals, B97\cite{B97} and B97-3\cite{B97_3}. B97 and B97-3 reached $0.55$ and $0.68~\rm kcal/mol$ MAE respectively when corrected with MBD. 
Both had MAEs $>3~\rm kcal/mol$ prior to correction with MBD, resulting in an impressive fractional improvement of $\Delta = 5.7$. B97 and B97-3 have among the shortest range exchange, resulting in small $\beta_{\rm opt}$ values of $0.50$ and $0.45$ respectively. The final functional tested using a B97-style series expansion of exchange, was HCTH/407, a heavily parameterized GGA functional\cite{HCTH07}~\cite{HCTH07}. HCTH/407 performed the best of the three B97-style functionals with an MAE of $2.21~\rm kcal/mol$, but receives the smallest benefit from MBD, only lowering the MAE to $0.73~\rm kcal/mol$. 

\begin{figure}[t!]
\centering
\includegraphics[width=228pt]{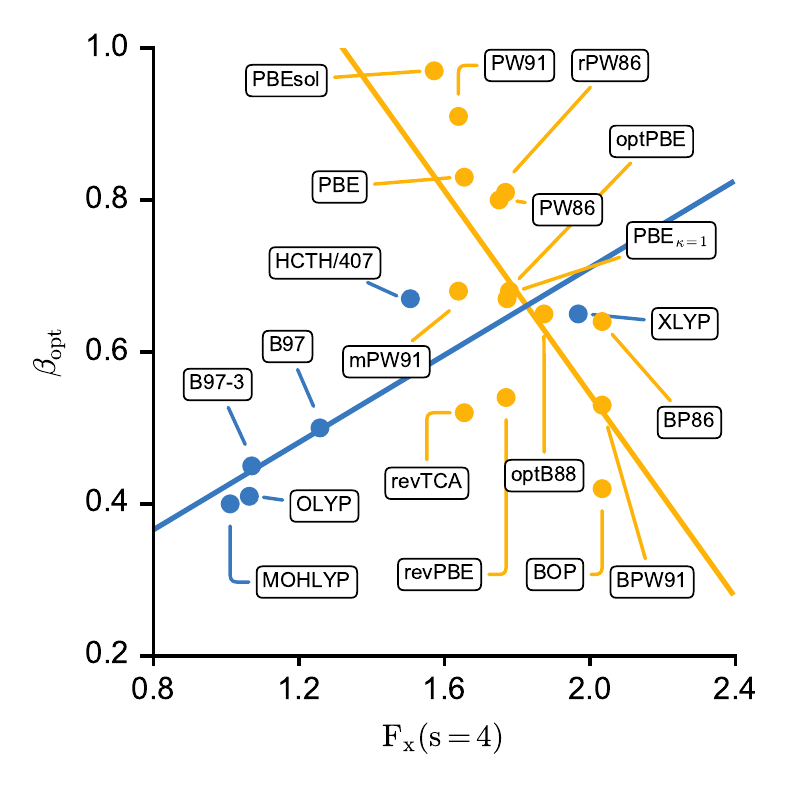}
\caption{The GGA enhancement factor evaluated at a reduced gradient of $s=4$ plotted against the associated optimal MBD range-separation parameter, $\beta_{\rm opt}$. There are two broad groups of GGA functionals, with the first, shown in yellow, corresponding to exchange functionals with PBE style exchange, and the second, shown in blue, corresponding to those that with B97 style exchange. 
\label{fig:en_fact}}
\end{figure}

TPSS\cite{TPSS_ref1,TPSS_ref2} was found to be one of the best performing functionals in our test set when corraected with MBD, with a total MAE of $0.44~\rm kcal/mol$, which is only $0.07~\rm kcal/mol$ above PBE+MBD -- an effectively negligible difference in overall performance. Split by category, we observe that TPSS+MBD performs among the best for dimers with hydrogen-bonded and ``other'' binding motifs, and acceptably well for dispersion-bound systems. MBD yields a remarkable 1200\% improvement ($\Delta=12$) for TPSS on hydrogen-bonded systems. We also observe very little spread in the $\beta_{\rm opt}$ among the different classes of interaction motif. Because of this consistency in both $\beta_{\rm opt}$ and accuracy between various classes of interactions, we believe that among the functionals tested, TPSS is the optimal one to be paired with MBD. 


Throughout this discussion we have divided functionals according to the classification of the exchange functional because of the interplay between the repulsive exchange wall and the range-separation of the long-range correlation correction. This analysis is made more explicit by examining the behavior of the gradient enhancement factor of the exchange functional. In Figure \ref{fig:en_fact} we plot the gradient enhancement factor $F_x$ evaluated at a reduced gradient, $s=4$, against $\beta_{\rm opt}$ for each functional. We have chosen $s=4$ because this is close to the value of the reduced gradient in the intermolecular range where noncovalent interactions tend to dominate. We find that for the two different classes of gradient enhancement factor present (B97/OPTX-like and PBE-like), a linear fit determines near optimal $\beta$s. $\beta$ values determined in this manner are hereafter referred to as $\beta_{\rm fit}$. We have evaluated the MAE for MBE@$\beta_{\rm fit}$ for all GGA and global hybrid functionals in this study, and presented these results in green in Figure \ref{fig:maebars}. Although the performance of MBD with an optimize $\beta$ is superior, if optimizing $\beta$ for your functional of choice is  not an option, we have found that we are able to obtain much of the benefits of MBD using a $\beta_{\rm fit}$ value. In many cases, we are able to reach chemical accuracy through this fitting process. We attribute much of this success to previously discussed the flat minima observed in the MAE curve when $\beta$ is varied. Indeed, the MAE across the S66$\times$8 set typically only changes by $0.2~\rm kcal/mol$ for as much as a $15\%$ change in $\beta$ away from the minimum. 

Overall, we find that the MBD model is able to compensate for the deficiencies of a variety of GGA and global hybrid functionals, with regard to their ability to describe noncovalent interaction energies. When the range-separation parameter is optimally tuned, many different functionals are able to achieve MAEs on the S66$\times$8 set that are well below $1~\rm kcal/mol$. Given the flat minima and the diversity of the S66 test set, we expect that the optimal $\beta$ determined in this study will be transferable to a wide range of systems. Additionally, the general functional form that we have determined for $\beta_{\rm fit}$, and the consistently high quality results that MBD@$\beta_{\rm fit}$ achieves, provides us with a way forward to applying MBD to a broader range of DFT functionals. This is an encouraging result, given that this study has only considered a small subset of existing functionals.

\section{Conclusions}
By determining the optimal value of the MBD range-separation parameter, $\beta$ for a range of functionals, we have enabled the further adoption of MBD as a highly accurate, low computational cost, correction for noncovalent interactions. Indeed, this work helps to ameliorate the common criticism that MBD is only applicable in conjunction with PBE or PBE0\cite{grimme_chemrev_mf}. We have shown that with the optimal range-separation of MBD, most functionals can obtain chemical accuracy for the interaction energy on the S66$\times$8 set. This provides a way forward towards more accurate predictions of the structure, energetics, and function of many systems previously inaccessible with uncorrected DFT. 

While we have only tested a small subset of available GGAs, our linear equation provides a near-optimal $\beta$ parameter. In most cases, the MAE for functionals with the parameterized $\beta$ were still able to achieve an accuracy better than $1~\rm kcal/mol$. We recommend finding the optimal $\beta$ value through MAE minimization, but recognize that this may not be possible in all cases. For those instances, we have shown that parameterized $\beta$ values will provide suitable results.

While we have provided reliable fits for twenty four different functionals, and many of them were developed to treat specific classes of molecules or properties, we recommend using either PBE+MBD or TPSS+MBD as general purpose functionals. Not only are PBE and TPSS two of the most popular and well understood functionals with sound non-empirical derivations, we observed mean absolute errors of $0.37~\rm kcal/mol$ and $0.44~\rm kcal/mol$ with respect to S66$\times$8 the CCSD(T)/CBS references for PBE+MBD and TPSS+MBD respectively. Additionally, their optimal range-separation values do not significantly depend on the type of noncovalent interaction, suggesting that the MBD correction to both functionals is quite transferable.

Future work will determine the optimal range separation parameter for a wider range of commonly used hybrids, meta-hybrids, and double hybrid functionals. Once a broader class of functionals has been parameterized for use with MBD, it will be interesting to see how the self-consistent application of MBD performs on the recently released 3B-69 benchmark set of three-body interactions\cite{3b69}, which have called into question whether many-body induction is at the root of many errors that were previously ascribed to dispersion. Because MBD can be applied self-consistently, it would also be interesting to explore the effect of the MBD contribution to the Kohn-Sham potential on the hole sum-rule. Such an analysis could provide new insights in understanding the underlying physics that results from the inclusion of many-body dispersion corrections to DFT.

\begin{acknowledgement}
The authors thank Narbe Mardirossian, Robert A. DiStasio Jr. and Alexandre Tkatchenko for useful discussions.
This research used resources of the Odyssey computing cluster supported by the FAS Division of Science, Research Computing Group at Harvard University.
T.M. acknowledges support from the NSF Graduate Research Fellowship Program.  
M.B.-F. acknowledges support from the DOE Office of Science Graduate Fellowship Program, made possible in part by the American Recovery and Reinvestment Act of 2009, administered by ORISE-ORAU under Contract No. DE-AC05-06OR23100. 
D.R. acknowledges support from NVIDA - CCOE Higher Education and Research Award.
%
%
T.M., D.K. and A.A.-G. acknowledge support from the STC Center for Integrated Quantum Materials, NSF Grant No. DMR-1231319.
All opinions expressed in this paper are the authors' and do not necessarily reflect the policies and views of DOE, ORAU, ORISE, or NSF.
\end{acknowledgement}

\bibliography{bibliography.bib}

\onecolumn
\section{Supporting Information}

\section{S22 benchmark set}

\FloatBarrier

\begin{table}[!htbp]
\captionsetup{width=0.5\textwidth}
\caption{Mean absolute error (MAE) in kcal/mol with respect to Takatani's revised S22 benchmark set~\cite{jurecka_s22_2006,takatani_s22a_2010} with the bare exchange-correlation functional and with the optimized range-separated MBD correction. $\beta_{\rm opt}$ is the optimized MBD range-separation parameter.\label{tab:s22mae}}
\begin{tabular}{ l ccc }
\toprule
\captionsetup{width=.5\textwidth}
 & \multicolumn{3}{c}{ MAE (kcal/mol) }  \\ \cmidrule(r){2-4} 
 Functional & Bare & MBD & $\beta_{\rm opt}$   \\
\midrule
B97-3 & 7.055 & 0.828 & 0.42\\
B97    & 5.766 & 0.768 & 0.48\\
TPSS & 4.75 & 0.801 & 0.55\\
MOHLYP & 8.15 & 2.921 & 0.42\\
OLYP & 7.124 & 2.675 & 0.49\\
BOP & 6.628 & 2.308 & 0.51\\
BPW91 	& 5.097 & 1.614 & 0.63\\
revTCA 	& 5.633 & 1.605 & 0.54\\
revPBE 	& 5.111 & 1.347 & 0.59\\
XLYP 	& 4.372 & 1.053 & 0.67\\
BP86 	& 3.948 & 1.109 & 0.71\\
optB88 	& 3.735 & 1.213 & 0.70\\
optPBE 	& 3.696 & 0.902 & 0.71\\
PBE$_{\kappa=1}$ & 3.772 & 0.743 & 0.70\\
mPW91 	& 3.675 & 0.895 & 0.71\\
AM05 	& 3.37 & 1.056 & 0.72\\
HCTH/407 & 3.949 & 0.828 & 0.64\\
PW86 	& 2.782 & 0.506 & 0.78\\
rPW86 	& 2.722 & 0.526 & 0.78\\
PBE 		& 2.516 & 0.601 & 0.81\\
PW91 	& 2.232 & 0.71 & 0.86\\
PBEsol 	& 1.898 & 0.968 & 0.95\\
SVWN 	& 2.323 & 3.068 & 1.19$^*$\\
SPZ 		& 2.065 & 2.789 & 1.19$^*$\\
\bottomrule
\multicolumn{4}{p{2.5in}}{$^*$ Maximum value of $\beta$ considered. LDA functionals are expected to optimize to $\beta \to \infty$.}
\end{tabular}
\end{table}

\begin{table}[!htbp]
\captionsetup{width=0.5\textwidth}
\caption{Mean absolute relative percentage error (MARE) with respect to Takatani's revised S22 benchmark set~\cite{jurecka_s22_2006,takatani_s22a_2010} with the bare exchange-correlation functional and with the optimized range-separated MBD correction. $\beta_{\rm opt}$ is the optimized MBD range-separation parameter.\label{tab:s22mare}}
\begin{tabular}{ l ccc }
\toprule
\captionsetup{width=.5\textwidth}
 & \multicolumn{3}{c}{ MARE (\%) }  \\ \cmidrule(r){2-4} 
 Functional & Bare & MBD & $\beta_{\rm opt}$   \\
\midrule
B97-3			&  124	&	18	&	0.42 \\
B97				&  102	&	18	&	0.49 \\
TPSS			&	96	&	17	&	0.51 \\
MOHLYP			&  167	&	51	&	0.38 \\
OLYP			&  150	&	47	&	0.39 \\
BOP				&  146	&	42	&	0.39 \\
BPW91			&  119	&	35	&	0.42 \\
revTCA			&  115	&	31	&	0.42 \\
revPBE			&  109	&	27	&	0.42 \\
XLYP			&	95	&	25	&	0.58 \\
BP86			&	96	&	26	&	0.57 \\
optB88			&	92	&	25	&	0.52 \\
optPBE			&	84	&	18	&	0.62 \\
PBE$_{\kappa=1}$&	81	&	15	&	0.64 \\
mPW91			&	84	&	18	&	0.62 \\
AM05			&	84	&	20	&	0.66 \\
HCTH/407		&	72	&	22	&	0.66 \\
PW86			&	63	&	 9	&	0.75 \\
rPW86			&	61	&	 9	&	0.76 \\
PBE				&	56	&	10	&	0.81 \\
PW91			&	47	&	14	&	0.86 \\
PBEsol			&	41	&	11	&	0.95 \\
SVWN		&	35	&	52	&	1.19$^*$ \\
SPZ		&	31	&	47	&	1.19$^*$ \\
\bottomrule
\multicolumn{4}{p{2.5in}}{$^*$ Maximum value of $\beta$ considered. LDA functionals are expected to optimize to $\beta \to \infty$.}
\end{tabular}
\end{table}

\cleardoublepage
\section{S66 benchmark set}
\FloatBarrier
\begin{table}[!htbp]
\captionsetup{width=0.5\textwidth}
\caption{Mean absolute error (MAE) in kcal/mol with respect to the S66 benchmark set~\cite{rezac_s66_2011} with the bare exchange-correlation functional and with the optimized range-separated MBD correction. $\beta_{\rm opt}$ is the optimized MBD range-separation parameter.\label{tab:s66mae}}
\begin{tabular}{ l ccc }
\toprule
\captionsetup{width=.5\textwidth}
 & \multicolumn{3}{c}{ MAE (kcal/mol) }  \\ \cmidrule(r){2-4} 
 Functional & Bare & MBD & $\beta_{\rm opt}$   \\
\midrule
B97-3			&	3.86	&	0.68	&	0.45		\\
B97				&	3.14	&	0.55	&	0.50		\\
TPSS			&	2.90	&	0.44	&	0.56		\\
MOHLYP			&	4.86	&	1.34	&	0.40		\\
OLYP			&	4.29	&	1.21	&	0.41		\\
BOP				&	4.15	&	1.15	&	0.42		\\
BPW91			&	3.33	&	1.02	&	0.53		\\
revTCA			&	3.31	&	0.81	&	0.52		\\
revPBE			&	3.13	&	0.71	&	0.54		\\
XLYP			&	2.63	&	0.61	&	0.65		\\
BP86			&	2.62	&	0.72	&	0.64		\\
optB88			&	2.55	&	0.75	&	0.65		\\
optPBE			&	2.34	&	0.49	&	0.67		\\
PBE$_\kappa=1$	&	2.31	&	0.42	&	0.68		\\
mPW91			&	2.30	&	0.49	&	0.68		\\
AM05			&	2.25	&	0.64	&	0.71		\\
HCTC/407		&	2.21	&	0.73	&	0.67		\\
PW86			&	1.76	&	0.38	&	0.80		\\
rPW86			&	1.71	&	0.39	&	0.81		\\
PBE				&	1.57	&	0.37	&	0.83		\\
PW91			&	1.25	&	0.50	&	0.91		\\
PBEsol			&	1.23	&	0.55	&	0.97		\\
SVWN			&	1.49	&	1.88	&	1.19$^*$	\\	
SPZ		  		&	1.37	&	1.76	&	1.19$^*$	\\
\bottomrule
\multicolumn{4}{p{2.5in}}{$^*$ Maximum value of $\beta$ considered. LDA functionals are expected to optimize to $\beta \to \infty$.}
\end{tabular}
\end{table}

\begin{table}[!htbp]
\captionsetup{width=.5\textwidth}
\caption{Mean absolute relative percentage error (MARE) with respect to the S66 benchmark set~\cite{rezac_s66_2011} with the bare exchange-correlation functional and with the optimized range-separated MBD correction. $\beta_{\rm opt}$ is the optimized MBD range-separation parameter.\label{tab:s66mare}}
\begin{tabular}{ l ccc }
\toprule
 & \multicolumn{3}{c}{ MARE (\%) }  \\ \cmidrule(r){2-4} 
 Functional & Bare & MBD & $\beta_{\rm opt}$   \\
\midrule
B97-3			&	146	&	47	&	0.46 \\
B97				&	117	&	42	&	0.57 \\
TPSS			&	117	&	28	&	0.53 \\
MOHLYP			&	185	&	62	&	0.39 \\
OLYP			&	163	&	58	&	0.40 \\
BOP				&	161	&	45	&	0.40 \\
BPW91			&	137	&	35	&	0.43 \\
revTCA			&	127	&	43	&	0.47 \\
revPBE			&	118	&	31	&	0.46 \\
XLYP			&	112	&	30	&	0.57 \\
BP86			&	110	&	30	&	0.55 \\
optB88			&	113	&	33	&	0.55 \\
optPBE			&	105	&	31	&	0.62 \\
PBE$_\kappa=1$	&	103	&	30	&	0.65 \\
mPW91			&	 94	&	21	&	0.63 \\
AM05			&	104	&	31	&	0.66 \\
HCTH/407		&	169	&  117	&	0.65 \\
PW86			&	 80	&	22	&	0.76 \\
rPW86			&	 80	&	24	&	0.76 \\
PBE				&	 76	&	23	&	0.80 \\
PW91			&	 58	&	30	&	0.88 \\
PBEsol			&	 55	&	19	&	0.95 \\
SVWN		&	 44	&	54	&	1.19$^*$ \\
SPZ		&	 42	&	49	&	1.19$^*$ \\
\bottomrule
\multicolumn{4}{p{2.5in}}{$^*$ Maximum value of $\beta$ considered. LDA functionals are expected to optimize to $\beta \to \infty$.}
\end{tabular}
\end{table}

\cleardoublepage
\section{S66$\times$8 benchmark set}

\begin{table}[H]
\captionsetup{width=.5\textwidth}
\caption{Mean absolute error (MAE) in kcal/mol with respect to the S66$\times$8 benchmark set~\cite{rezac_s66x8_2011} with the bare exchange-correlation functional and with the optimized range-separated MBD correction. $\beta_{\rm opt}$ is the optimized MBD range-separation parameter. \label{tab:s66x8mae}}
\begin{tabular}{ l  ccc }
\toprule
 \multicolumn{1}{l}{Functional} &  \multicolumn{1}{c}{Bare} & \multicolumn{1}{c}{MBD} & \multicolumn{1}{c}{$\beta_{\rm opt}$}  \\
\midrule
B97-3            & 3.86 & 0.68 & 0.45     \\
B97              & 3.14 & 0.55 & 0.50     \\
TPSS             & 2.90 & 0.44 & 0.56     \\
MOHLYP           & 4.86 & 1.34 & 0.40     \\
OLYP             & 4.29 & 1.21 & 0.41     \\
BOP              & 4.15 & 1.15 & 0.42     \\
BPW91            & 3.33 & 1.02 & 0.53     \\
revTCA           & 3.31 & 0.81 & 0.52     \\
revPBE           & 3.13 & 0.71 & 0.54     \\
XLYP             & 2.63 & 0.61 & 0.65     \\
BP86             & 2.62 & 0.72 & 0.64     \\
optB88           & 2.55 & 0.75 & 0.65     \\
optPBE           & 2.34 & 0.49 & 0.67     \\
PBE$_{\kappa=1}$ & 2.31 & 0.42 & 0.68     \\
mPW91            & 2.30 & 0.49 & 0.68     \\
AM05             & 2.25 & 0.64 & 0.71     \\
HCTC/407         & 2.21 & 0.73 & 0.67     \\
PW86             & 1.76 & 0.38 & 0.80     \\
rPW86            & 1.71 & 0.39 & 0.81     \\
PBE              & 1.57 & 0.37 & 0.83     \\
PW91             & 1.25 & 0.50 & 0.91     \\
PBEsol           & 1.23 & 0.55 & 0.97     \\
SVWN        	 & 1.49 & 1.88 & 1.19$^*$ \\
SPZ         	 & 1.37 & 1.76 & 1.19$^*$ \\
\bottomrule
\multicolumn{4}{p{2.5in}}{$^*$ Maximum value of $\beta$ considered. LDA functionals are expected to optimize to $\beta \to \infty$.}
\end{tabular}
\end{table}

\begin{table}[H]
\captionsetup{width=.5\textwidth}
\caption{Mean absolute relative percentage error (MARE) with respect to the S66$\times$8 benchmark set~\cite{rezac_s66x8_2011} with the bare exchange-correlation functional and with the optimized range-separated MBD correction. $\beta_{\rm opt}$ is the optimized MBD range-separation parameter. \label{tab:s66x8mare}}
\begin{tabular}{ l ccc }
\toprule
 & \multicolumn{3}{c}{ MARE (\%) }  \\ \cmidrule(r){2-4} 
 Functional & Bare & MBD & $\beta_{\rm opt}$   \\
\midrule
B97-3            & 146 &  47 & 0.46     \\
B97              & 116 &  42 & 0.57     \\
TPSS             & 117 &  28 & 0.53     \\
MOHLYP           & 185 &  62 & 0.39     \\
OLYP             & 163 &  58 & 0.40     \\
BOP              & 161 &  44 & 0.40     \\
BPW91            & 137 &  35 & 0.43     \\
revTCA           & 127 &  43 & 0.47     \\
revPBE           & 118 &  31 & 0.46     \\ 
XLYP             & 112 &  30 & 0.57     \\
BP86             & 110 &  30 & 0.55     \\
optB88           & 113 &  33 & 0.55     \\
optPBE           & 105 &  31 & 0.62     \\
PBE$_{\kappa=1}$ & 103 &  30 & 0.65     \\
mPW91            &  94 &  21 & 0.63     \\
AM05             & 104 &  31 & 0.66     \\
HCTH/407         & 169 & 117 & 0.65     \\
PW86             &  80 &  22 & 0.76     \\
rPW86            &  80 &  23 & 0.76     \\
PBE              &  76 &  23 & 0.80     \\
PW91             &  58 &  30 & 0.88     \\
PBEsol           &  55 &  19 & 0.95     \\
SVWN        	 &  44 &  54 & 1.19$^*$ \\
SPZ         	 &  42 &  49 & 1.19$^*$ \\
\bottomrule
\multicolumn{4}{p{2.5in}}{$^*$ Maximum value of $\beta$ considered. LDA functionals are expected to optimize to $\beta \to \infty$.}
\end{tabular}
\end{table}

\begin{table}[H]
\caption{Mean absolute error (MAE) in kcal/mol with respect to subsets of the S66$\times$8 benchmark set~\cite{rezac_s66x8_2011} categorized by bonding motif, with the bare exchange-correlation functional and with the optimized range-separated MBD correction. $\beta_{\rm opt}$ is the optimized MBD range-separation parameter.\label{tab:s66x8mae_broken_down}}
\begin{tabular}{ l |  ccc | ccc | ccc }
\toprule
\multicolumn{1}{l}{} & \multicolumn{3}{c}{ H-Bonded } & \multicolumn{3}{c}{ Dispersion } & \multicolumn{3}{c}{ Other } \\ 
 \cmidrule(r){2-4}  \cmidrule(r){5-7}  \cmidrule(r){8-10} 
  \multicolumn{1}{l}{Functional} &  \multicolumn{1}{c}{Bare} & \multicolumn{1}{c}{MBD} & \multicolumn{1}{c}{$\beta_{\rm opt}$} &  \multicolumn{1}{c}{Bare} & \multicolumn{1}{c}{MBD} & \multicolumn{1}{c}{$\beta_{\rm opt}$} & \multicolumn{1}{c}{Bare} & \multicolumn{1}{c}{MBD} & \multicolumn{1}{c}{$\beta_{\rm opt}$} \\
\midrule
B97-3            & 3.93 & 0.36 & 0.44     & 4.42 & 0.99 & 0.55     & 3.13 & 0.51 & 0.42 \\
B97               & 3.00 & 0.35 & 0.49     & 3.80 & 0.80 & 0.61     & 2.54 & 0.33 & 0.52 \\
TPSS            & 2.18 & 0.18 & 0.57     & 4.07 & 0.68 & 0.48     & 2.39 & 0.30 & 0.49 \\
MOHLYP       & 3.88 & 0.86 & 0.45     & 6.52 & 1.70 & 0.39     & 4.09 & 0.87 & 0.39 \\
OLYP             & 3.10 & 0.62 & 0.49     & 5.97 & 1.45 & 0.41     & 3.73 & 0.76 & 0.39 \\
BOP               & 2.98 & 0.64 & 0.52     & 5.77 & 1.31 & 0.40     & 3.64 & 0.61 & 0.40 \\
BPW91          & 2.08 & 0.58 & 0.62     & 4.88 & 0.77 & 0.43     & 2.99 & 0.52 & 0.42 \\
revTCA           & 2.42 & 0.30 & 0.55     & 4.61 & 1.01 & 0.43     & 2.83 & 0.50 & 0.42 \\
revPBE           & 2.17 & 0.33 & 0.59     & 4.44 & 0.78 & 0.43     & 2.72 & 0.38 & 0.44 \\
XLYP              & 1.45 & 0.19 & 0.71     & 4.06 & 0.81 & 0.49     & 2.34 & 0.36 & 0.52 \\
BP86              & 1.38 & 0.45 & 0.75     & 4.04 & 0.74 & 0.47     & 2.41 & 0.48 & 0.53 \\
optB88            & 1.27 & 0.51 & 0.79     & 4.03 & 0.76 & 0.49     & 2.33 & 0.51 & 0.57 \\
optPBE           & 1.29 & 0.35 & 0.75     & 3.60 & 0.54 & 0.59     & 2.11 & 0.32 & 0.62 \\
PBE$_{\kappa=1}$ & 1.33 & 0.26 & 0.74     & 3.51 & 0.55 & 0.60     & 2.06 & 0.26 & 0.63 \\
mPW91           & 1.21 & 0.31 & 0.77     & 3.60 & 0.53 & 0.57     & 2.06 & 0.29 & 0.64 \\
AM05              & 0.94 & 0.46 & 0.87     & 3.69 & 0.65 & 0.57     & 2.09 & 0.39 & 0.68 \\
HCTC/407       & 1.24 & 0.47 & 0.70     & 3.47 & 1.10 & 0.60     & 1.88 & 0.52 & 0.63 \\
PW86              & 0.71 & 0.21 & 0.93     & 2.87 & 0.35 & 0.74     & 1.70 & 0.27 & 0.76 \\
rPW86             & 0.66 & 0.21 & 0.95     & 2.81 & 0.33 & 0.75     & 1.66 & 0.27 & 0.77 \\
PBE                 & 0.60 & 0.31 & 0.99     & 2.66 & 0.32 & 0.77     & 1.44 & 0.19 & 0.82 \\
PW91              & 0.43 & 0.36 & 1.19$^*$ & 2.23 & 0.45 & 0.83     & 1.06 & 0.32 & 0.91 \\
PBEsol            & 0.70 & 0.74 & 1.19$^*$ & 1.94 & 0.32 & 0.93     & 1.03 & 0.26 & 0.96 \\
SVWN             & 2.66 & 2.93 & 1.19$^*$ & 0.82 & 1.35 & 1.19$^*$ & 0.93 & 1.28 & 1.19$^*$ \\
SPZ         	       & 2.49 & 2.76 & 1.19$^*$ & 0.72 & 1.25 & 1.19$^*$ & 0.83 & 1.18 & 1.19$^*$ \\
\bottomrule
\multicolumn{10}{p{5.5in}}{$^*$ Maximum value of $\beta$ considered. LDA functionals are expected to optimize to $\beta \to \infty$.}
\end{tabular}
\end{table}

\begin{table}[H]
\caption{Mean absolute relative percentage error (MARE) with respect to subsets of the S66$\times$8 benchmark set~\cite{rezac_s66x8_2011} categorized by bonding motif, with the bare exchange-correlation functional and with the optimized range-separated MBD correction. $\beta_{\rm opt}$ is the optimized MBD range-separation parameter.\label{tab:s66x8mare_broken_down}}
\begin{tabular}{ l | ccc | ccc | ccc }
\toprule
 \multicolumn{1}{c}{} & \multicolumn{3}{c}{ H-Bonded } & \multicolumn{3}{c}{ Dispersion } & \multicolumn{3}{c}{ Other } \\ 
 \cmidrule(r){2-4}  \cmidrule(r){5-7}  \cmidrule(r){8-10} 
 \multicolumn{1}{l}{Functional} &  \multicolumn{1}{c}{Bare} & \multicolumn{1}{c}{MBD} & \multicolumn{1}{c}{$\beta_{\rm opt}$} &  \multicolumn{1}{c}{Bare} & \multicolumn{1}{c}{MBD} & \multicolumn{1}{c}{$\beta_{\rm opt}$} & \multicolumn{1}{c}{Bare} & \multicolumn{1}{c}{MBD} & \multicolumn{1}{c}{$\beta_{\rm opt}$}   \\
\midrule
B97-3             & 51 &  7 & 0.45     & 265 & 100 & 0.55     & 118 & 28 & 0.42      \\
B97               & 39 &  7 & 0.50     & 215 &  89 & 0.63     &  92 & 20 & 0.53      \\
TPSS              & 31 &  3 & 0.57     & 229 &  60 & 0.45     &  89 & 16 & 0.50      \\
MOHLYP            & 55 & 13 & 0.41     & 350 & 127 & 0.39     & 144 & 38 & 0.39      \\
OLYP              & 45 & 11 & 0.47     & 308 & 119 & 0.40     & 132 & 34 & 0.39      \\
BOP               & 46 & 11 & 0.47     & 300 &  88 & 0.40     & 133 & 25 & 0.40      \\
BPW91             & 35 & 11 & 0.57     & 259 &  56 & 0.43     & 112 & 22 & 0.42      \\
revTCA            & 35 &  6 & 0.54     & 242 &  87 & 0.43     & 101 & 25 & 0.42      \\
revPBE            & 33 &  6 & 0.56     & 217 &  52 & 0.43     & 100 & 21 & 0.45      \\ 
XLYP              & 23 &  4 & 0.71     & 223 &  58 & 0.43     &  87 & 17 & 0.52      \\
BP86              & 25 &  9 & 0.71     & 210 &  48 & 0.46     &  91 & 22 & 0.50      \\
optB88            & 24 &  9 & 0.73     & 223 &  58 & 0.50     &  89 & 22 & 0.50      \\
optPBE            & 21 &  6 & 0.73     & 211 &  64 & 0.59     &  79 & 17 & 0.63      \\
PBE$_{\kappa=1}$  & 21 &  4 & 0.72     & 208 &  67 & 0.62     &  77 & 15 & 0.63      \\
mPW91             & 20 &  5 & 0.75     & 185 &  39 & 0.58     &  75 & 14 & 0.64      \\
AM05              & 18 &  7 & 0.81     & 209 &  59 & 0.55     &  81 & 20 & 0.68      \\
HCTH/407          & 18 & 11 & 0.76     & 397 & 285 & 0.60     &  81 & 40 & 0.64      \\
PW86              & 12 &  4 & 0.92     & 160 &  42 & 0.74     &  66 & 14 & 0.76      \\
rPW86             & 11 &  4 & 0.94     & 162 &  46 & 0.75     &  64 & 14 & 0.77      \\
PBE               & 11 &  5 & 0.96     & 160 &  46 & 0.75     &  56 & 11 & 0.82      \\
PW91              &  7 &  7 & 1.19$^*$ & 123 &  55 & 0.83     &  40 & 21 & 0.91      \\
PBEsol            & 10 &  9 & 1.19$^*$ & 107 &  27 & 0.94     &  45 & 14 & 0.96      \\
SVWN        	  & 35 & 39 & 1.19$^*$ &  58 &  76 & 1.19$^*$ &  39 & 45 & 1.19$^*$  \\
SPZ         	  & 33 & 37 & 1.19$^*$ &  57 &  68 & 1.19$^*$ &  35 & 42 & 1.19$^*$  \\
\bottomrule
\multicolumn{10}{p{5.5in}}{$^*$ Maximum value of $\beta$ considered. LDA functionals are expected to optimize to $\beta \to \infty$.}
\end{tabular}
\end{table}

\clearpage
\section{Glossary}
\begin{table}[H]
\caption{Definition of abbreviations referring to the exchange-correlation functionals considered herein.\label{tab:fun_ref}}
\begin{tabular}{p{1.in}  p{2.5in}  p{2.5in}}
\toprule
Abbreviation   & Exchange & Correlation \\
 \midrule
SVWN 	   & Slater exchange				& Vosko, Wilk, \& Nussair `80, (RPA)  \\
SPZ 	   & Slater exchange      			& Perdew \& Zunger `81 local				\\
\midrule
BOP 		   & Becke `88 	 	      			& B88 one parameter progressive \\
BP86           & Becke `88            			& Perdew `86                      \\
BPW91          & Becke `88            			& Perdew \& Wang `91              \\
PW91           & Perdew \& Wang `91 			& Perdew \& Wang `91              \\
mPW91 & modified PW91 (Adomo \& Barone)    &  Perdew \& Wang `91  \\
PBE            & Perdew, Burke \& Ernzerhof 	& Perdew, Burke \& Ernzerhof      \\
PBEsol         & PBE reparameterized for solids & PBE reparameterized for solids \\
revPBE         & revised PBE           			& Perdew, Burke \& Ernzerhof      \\
PBE$_{\kappa=1}$ & PBE reparametrized for vdW   & Perdew, Burke \& Ernzerhof      \\
optPBE    	   & PBE reparametrized for vdW     & Perdew, Burke \& Ernzerhof      \\
rPW86          & PW86 reparametrized for vdW	& Perdew, Burke \& Ernzerhof      \\
optB88     	   & B88 reparametrized for vdW     & Perdew, Burke \& Ernzerhof      \\
PW86           & Perdew \& Wang `86         	& Perdew, Burke \& Ernzerhof  \\
revTCA         & Hammer, Hansen \& Norskov 		& revised Tognetti, Cortona, Adamo   \\
AM05           & Armiento \& Mattsson `05		& Armiento \& Mattsson `05        \\
HCTH/407       & \multicolumn{2}{l}{Handy \textit{et al.}'s functional fit to 407 molecules}     \\
OLYP           & Handy \& Cohen OPTX `01 		& Lee, Yang \& Parr               \\
MOHLYP         & scaled Handy \& Cohen OPTX `01 & scaled Lee, Yang \& Parr        \\
\multirow{2}{*}{XLYP}& local: 100\% Slater  	& \multirow{2}{*}{Lee, Yang \& Parr}  \\
			   & nonlocal: $\rm 72.2\%\; B88 + 34.7\%\; PW91$ & 	 \\
\midrule
TPSS           & \multicolumn{2}{l}{Tao, Perdew, Staroverov \& Scuseria  meta-GGA} \\
\midrule
B97 		   & \multicolumn{2}{l}{Becke `97 global hybrid}\\
B97-3          & \multicolumn{2}{l}{B97 reparameterized by  Keal and Tozer `05} \\
\bottomrule
\end{tabular}
\end{table}
\clearpage

\end{document}